# Rdgai: Classifying transcriptional changes using Large Language Models with a test case from an Arabic Gospel tradition

Robert Turnbull
Melbourne Data Analytics Platform
The University of Melbourne
Grattan St. Carlton VIC, Australia 3053
robert.turnbull@unimelb.edu.au
ORCID: 0000-0003-1274-6750

## Abstract

Application of phylogenetic methods to textual traditions has traditionally treated all changes as equivalent even though it is widely recognized that certain types of variants were more likely to be introduced than others. While it is possible to give weights to certain changes using a maximum parsimony evaluation criterion, it is difficult to state *a priori* what these weights should be. Probabilistic methods, such as Bayesian phylogenetics, allow users to create categories of changes, and the transition rates for each category can be estimated as part of the analysis. This classification of types of changes in readings also allows for inspecting the probability of these categories across each branch in the resulting trees. However, classification of readings is time-consuming, as it requires categorizing each reading against every other reading at each variation unit, presenting a significant barrier to entry for this kind of analysis. This paper presents Rdgai, a software package that automates this classification task using multi-lingual large language models (LLMs). The tool allows users to easily manually classify changes in readings and then it uses these annotations in the prompt for an LLM to automatically classify the remaining reading transitions. These classifications are stored in TEI XML and ready for downstream phylogenetic analysis. This paper demonstrates the application with data an Arabic translation of the Gospels.

## Introduction

Not all textual variants were created equal. Some types of changes in variant readings, such as orthographic changes, may have occurred at many independent times across a tradition. Other changes are so substantial that they were introduced only on a single occasion. The application of phylogenetic methods to textual traditions has traditionally treated all changes as equivalent. If the former types of variation which occurred frequently in the copying process are weighted the same as the strongly informative readings, then we risk drowning the true phylogenetic signal with noise.[1] Worse, we may introduce systematic bias which distorts the inference of the relationships between the textual witnesses. For this reason, types of variation deemed insignificant have often been filtered out of phylogenetic analysis. In textual traditions with limited evidence, this can dramatically reduce the number of available variation units. It is also difficult for the textual critic to know *a priori* exactly which variants to exclude. After the filtering process, the readings that remain are typically given the same weight in evaluating prospective phylogenetic trees. While it is possible to give weights to certain changes using a maximum parsimony evaluation criterion, it is difficult to determine what these weights should be.[2] Probabilistic methods, such as Bayesian phylogenetics, allow users to create categories of changes, and the transition rates for each category can be estimated as part of the analysis.[3] This classification of types of changes in readings also allows for estimating the probability of each type of change across branches in the resulting trees. This allows us to explore the habits of scribes across the branches of the tradition.

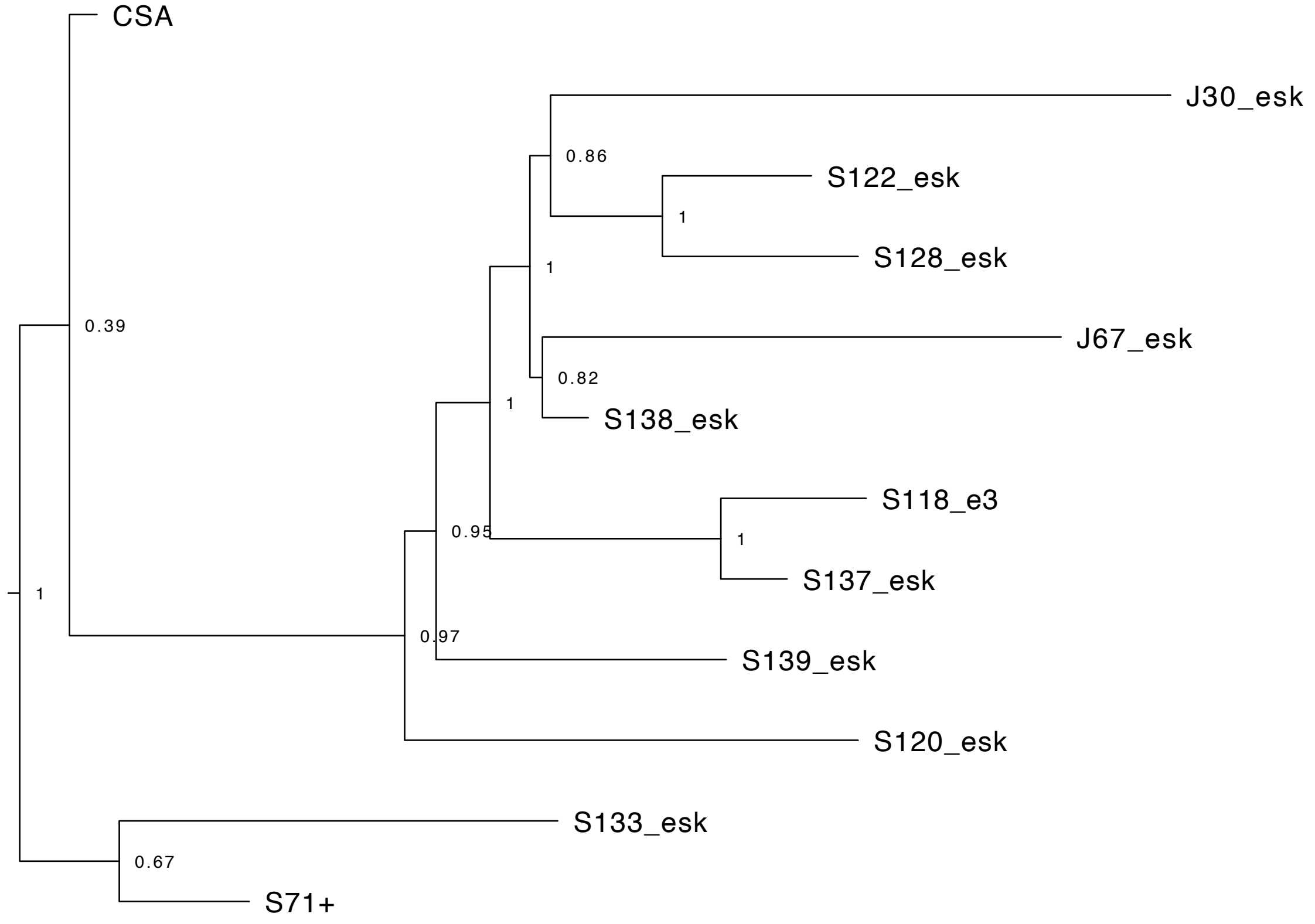

*Fig. 1. The maximum clade credibility tree of ar$^b$ with different rates for 'significant' changes and 'insignificant' ones.*

For example, Codex Sinaiticus Arabicus and its family preserve a translation of the New Testament Gospels from Greek into Arabic called ar$^b$.[4] This family includes three continuous text

---

[1] Trovato, *Everything You Always Wanted to Know about Lachmann's Method*, 112, 115, 184, 194, 216–17.

[2] Felsenstein, *Inferring Phylogenies*, 82–83.

[3] McCollum and Turnbull, "Using Bayesian Phylogenetics to Infer Manuscript Transmission History."

[4] Turnbull, *Codex Sinaiticus Arabicus and Its Family,* 176–204.

Gospel manuscripts and over ten lectionaries. In a previous work, I discuss creating a collation of the text of this family for John 5:36–8:51. This includes over 800 variation units and over 2,000 variant readings. Approximately 60 categories of change were used. These categories were divided into two classes: significant changes which affected the meaning and insignificant ones. These two classes of change were allowed to have different rates in the phylogenetic analysis. The resulting maximum clade credibility tree is shown in Fig. 1. It was found that insignificant changes occurred approximately 1.8 times that of significant changes. Using ancestral state reconstruction, the posterior probability distribution of different classes of changes across the branches of the tree could be inferred. For example, in the branch that led to Sinai ar. 137, there were mainly orthographic changes or additions of a conjunction but rarely changes involving a difference in meaning (Fig. 2). However in a branch leading to the archetype of a lectionary which used this translation, it was common for the scribe to substitute a word with a synonym (Fig. 3).

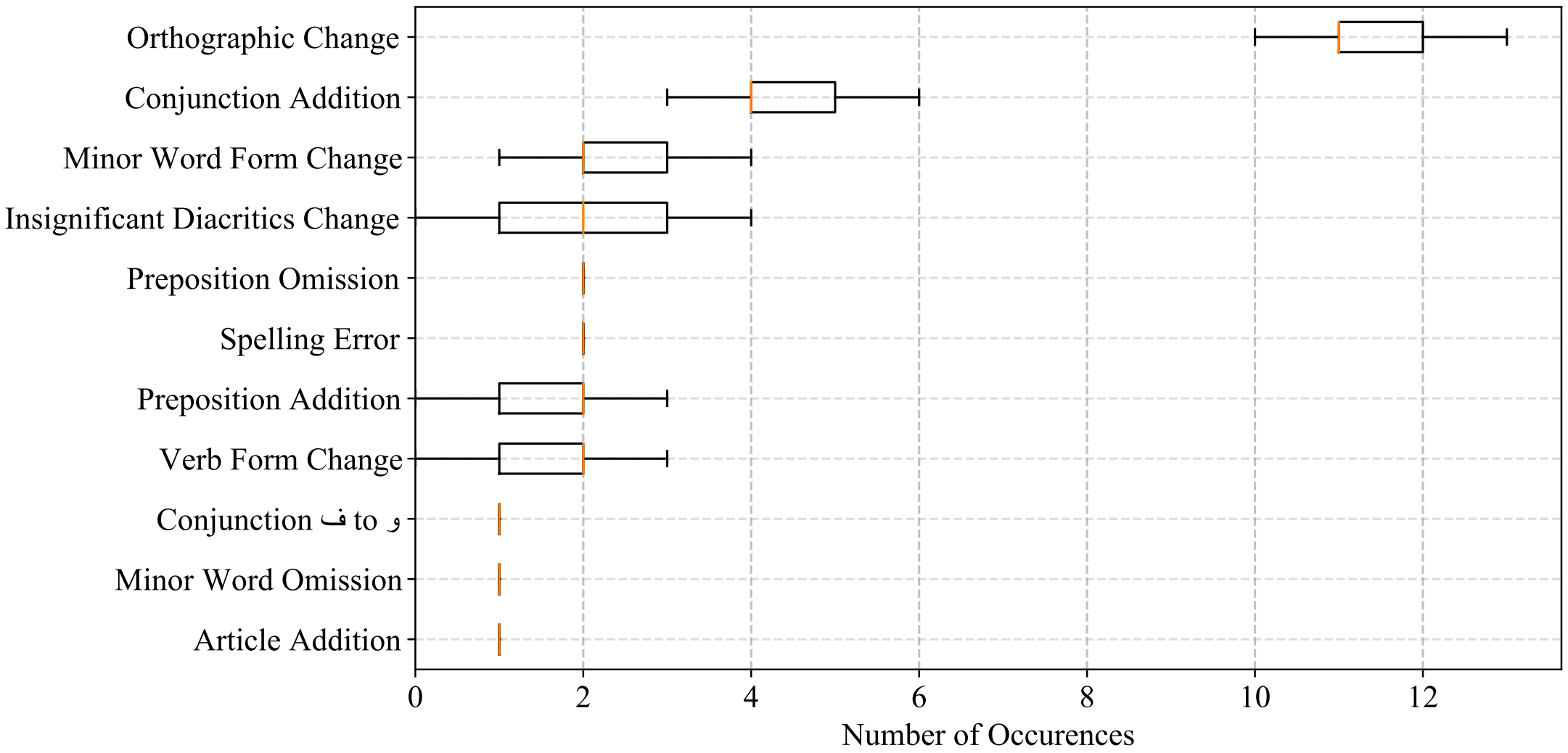

*Fig. 2. Posterior probability distribution of scribal changes in the branch leading to Sinai ar. 137.*

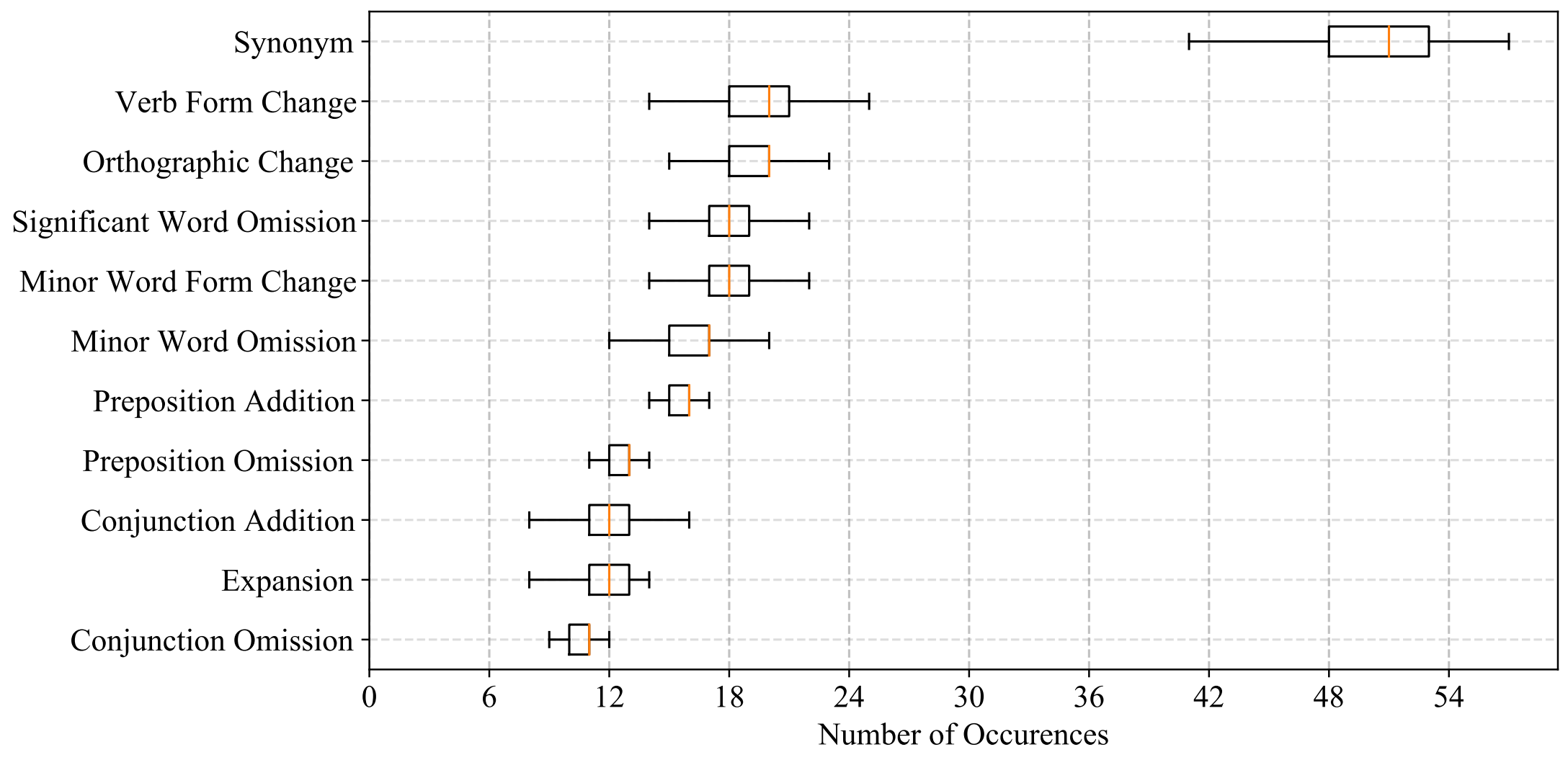

*Fig. 3. Posterior probability distribution of scribal changes in the branch leading to archetype of the esk2B lectionary.*

# Automated Classification

Classifying changes in variant readings not only improves the phylogenetic analysis itself by allowing different substitution rates, but also provides deeper insights into the transmission history. But this is a laborious process. It requires the manual classification of potentially many thousands of possible textual changes. This is a significant barrier to entry for this kind of study. Can this process be automated? One could define a list of programmatic rules involving regular expressions to compare the strings for both readings and classify the type of transition but arriving at these rules is extremely difficult; there will always be exceptions and the rules would only apply to a limited range of textual traditions and encodings.

A more promising avenue is the use of artificial intelligence to perform the classification task. One could train a neural network using supervised learning to take an annotated dataset of pairs of readings with a human classification of the type of transition between them. Given enough training data, such a network could potentially produce reasonable classifications but this model would only be applicable to the types of textual traditions and the categorization of reading transitions from the training dataset. The rise of Large Language Models (LLMs) provides an alternative avenue.

Language Models are artificial intelligence systems trained to predict the next token in a textual sequence.[5] The token could be a character, whole word or part of a word. It addresses the conditional probability problem of: 'given the sequence of text so far, what is the probability distribution of the next token?' This is an example of self-supervised learning—since the answer to that question is provided in the source text itself, no human annotator is required to provide the ground truth for the model to train. This allows models to be trained on the vast amounts of text available in the digital age. Models trained on vast textual corpora with a very large number of learnable weights/parameters are called Large Language Models or LLMs.

[5] Language Models can also be used to predict tokens within a sequence, but this article is focused on Causal Language Modelling for text generation.

The task of predicting the next token is very challenging. It requires the language model to learn a sophisticated internal representation of the context and structure within the textual sequence.[6] The sophisticated internal representations of language that LLMs generate mean that they are useful as pretrained models which can be fine-tuned on many other downstream tasks such as text classification, sentiment analysis and question clarification.[7] By training the model to predict the next token in a sequence, it is able to continuously generate new text. With enough capacity and training data, language models themselves can become an 'unsupervised multitask learner', as[8] state: 'high-capacity models trained to maximize the likelihood of a sufficiently varied text corpus begin to learn how to perform a surprising amount of tasks without the need for explicit supervision.' LLMs' performance in downstream tasks can be improved by including examples of the required task in the prompt context without needing to update the parameters of the model, leading to LLMs being described as 'Few-Shot Learners.'[9] This allows LLMs to be used for question answering with many Large Language Models being fine-tuned on curated datasets of questions and answers to produce 'Chat' or 'Instruct' models which are trained to answer questions as helpfully as possible. As the size (i.e. the number of weights/parameters) of state-of-the-art LLMs has grown, so too have the costs (both economic and environmental) of training them. It is unrealistic for individuals and smaller institutions to train these models and instead users can employ LLMs trained by others. Some models, such as OpenAI's ChatGPT and Anthropic's Claude, are available with a paid service. Others such as Meta's Llama are open-access and their weights available for download so that users can fine-tune the model for other tasks. To use an LLM to predict the classification of changes to readings, a custom dataset of readings with correct answers could be produced and an open-access LLM could be fine-tuned for that task. Alternatively, a description of the task with enough examples in the prompt could be provided. The latter is the approach used in this study.

One limitation of LLMs is known as 'hallucination'. Language models are simply optimized to predict the next token in a sequence. If the desired response to a prompt requires information that was available in the training data, then potentially this information has been retained within the parameters of the model (known as 'parametric knowledge') and the model can use this information to produce a response. But if the required information was not available in the training data or not at a sufficient frequency to be absorbed by the model, then the model will simply produce the most plausible text that it can generate. Chat models can be fine-tuned to reduce this with supervised learning and reinforcement learning from human-annotated data but fundamentally, LLMs were primarily rewarded for producing plausible probability distributions for future tokens. This means that LLMs frequently produce reasonably sounding results which simply are not connected to the real world. One way to mitigate this tendency is to include all required information as part of the prompt (known as 'contextual knowledge'). In this way, the LLM is used as a 'reasoning engine' rather than as a source of information in itself.

---

[6] Devlin et al., "BERT."

[7] Howard and Ruder, "Universal Language Model Fine-Tuning for Text Classification."

[8] Radford et al., "Language Models Are Unsupervised Multitask Learners."

[9] Brown et al., "Language Models Are Few-Shot Learners."

# Rdgai Software package

To facilitate the use of LLMs for classifying transitions between variant readings, we introduce Rdgai.[10] Rdgai takes as input a Text Encoding Initiative (TEI) XML file containing a critical apparatus.[11] First, the user can create a list of transition categories. The classification categories are defined in the <interpGrp type="transcriptional"> element in the header of the XML file. These categories can be edited and added to using the Rdgai graphical user interface to help users easily manipulate TEI XML files for a critical apparatus. It is helpful to also provide a description of each category in order to define it for other human readers of the XML and also the LLM.

Classification categories can come in reciprocal pairs so that if a transition from reading A to reading B is classified as category C, then the reverse transition from reading B to reading A can automatically be classed in the reciprocal category C'. In this case, categories C and C' can have their inverse in the corresp attribute of the interp element. Categories can also be symmetrical, such that if a transition from reading A to reading B is in symmetrical category S, then the same category can describe the transition between B and A. In this case, the category should not be given a corresp attribute.

## Manual Data Ingest

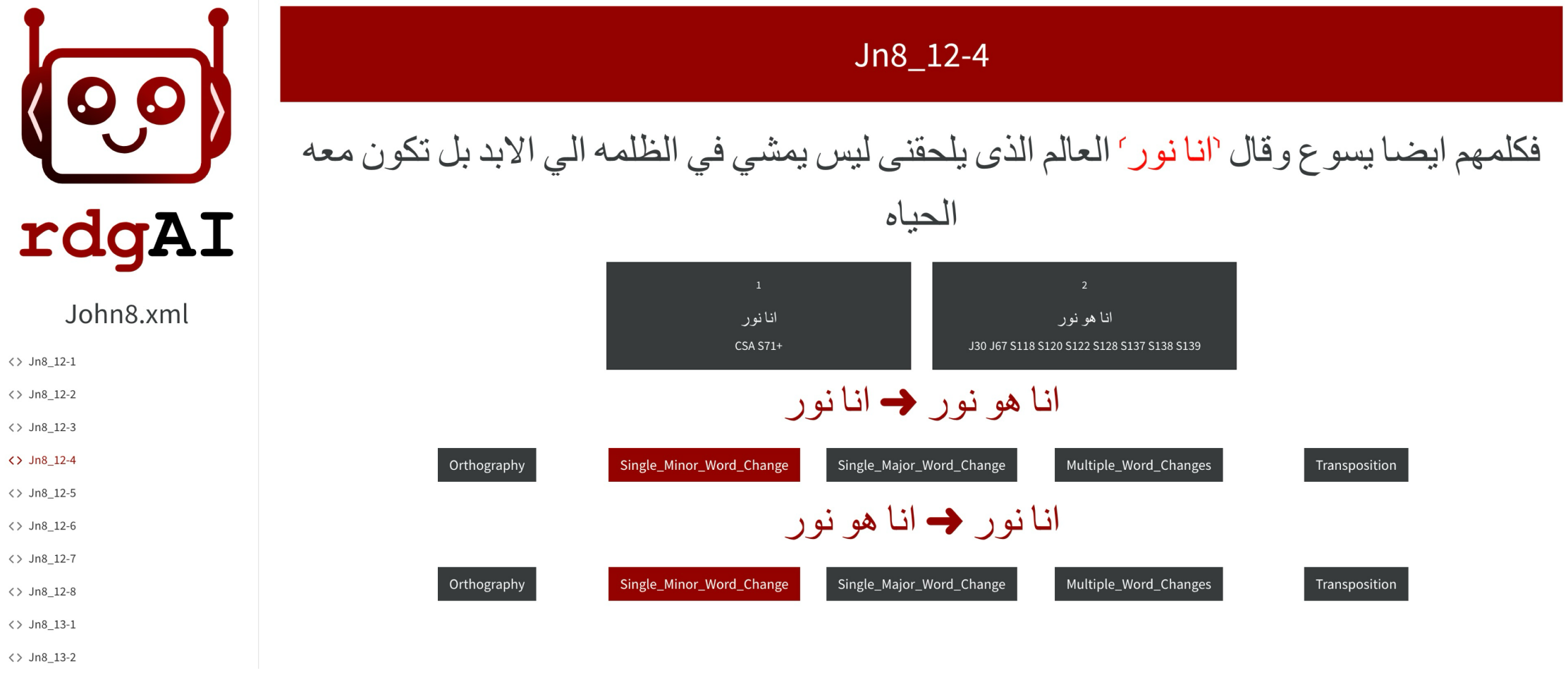

*Fig. 4. The Rdgai GUI for manually classifying textual transitions.*

[10] The name Rdgai derives from its function of classifying TEI <rdg> reading elements using AI.

[11] Ide and Sperberg-McQueen, "The TEI"; TEI Consortium, "TEI P5."

| | A | B | C | D | E | F | G | H |
|---|---|---|---|---|---|---|---|---|
| 1 | **App ID** | **Context** | **Active Reading ID** | **Passive Reading ID** | **Active Reading Text** | **Passive Reading Text** | **Description** | **Relation Type(s)** |
| 2 | Jn8_12-1 | فكلمهم ايضا يسوع وقال انا نور العالم T الذى يلحقنى ليس يمشي في الظلمه الي الابد بل تكون معه الحياه | 1 | 2 | | قال الرب للذين اتوا اليه من اليهود | Several words are added. | Multiple_Word_Changes |
| 3 | Jn8_12-1 | فكلمهم ايضا يسوع وقال انا نور العالم T الذى يلحقنى ليس يمشي في الظلمه الي الابد بل تكون معه الحياه | 1 | 3 | | قال الرب للذين اتو اليه من اليهود | Several words are added. | Multiple_Word_Changes |
| 4 | Jn8_12-1 | فكلمهم ايضا يسوع وقال انا نور العالم T الذى يلحقنى ليس يمشي في الظلمه الي الابد بل تكون معه الحياه | 1 | 4 | | قال الرب لليهود الذين اتوا اليه | Several words are added. | Multiple_Word_Changes |
| 5 | Jn8_12-1 | فكلمهم ايضا يسوع وقال انا نور العالم T الذى يلحقنى ليس يمشي في الظلمه الي الابد بل تكون معه الحياه | 2 | 3 | قال الرب للذين اتوا اليه من اليهود | قال الرب للذين اتو اليه من اليهود | The only change is the removal of the alif in اتوا which is merely an orthographic change. | Orthography |
| 6 | Jn8_12-1 | فكلمهم ايضا يسوع وقال انا نور العالم T الذى يلحقنى ليس يمشي في الظلمه الي الابد بل تكون معه الحياه | 2 | 4 | قال الرب للذين اتوا اليه من اليهود | قال الرب لليهود الذين اتوا اليه | More than one word is changed because by transposing اليهود it also affects the ل preposition. | Multiple_Word_Changes |

*Fig. 5. An Excel spreadsheet exported from rdgai.*

After defining the transition categories, the user can manually classify for transitions in readings using the rdgai Graphical User Interface (GUI) (Fig. 4). The user then clicks the appropriate button for the classification of each reading. The user can progress to the next variation unit by tapping the keyboard arrows.

There is also an option to export the readings and current classifications to Excel format (Fig. 5). This allows the user to edit classifications and add justifications whilst viewing the variation units in a tabular format. The files can be easily shared and edited in the cloud using Microsoft SharePoint. The Excel file can then be imported back into the TEI XML format using Rdgai.

## LLM Classification

Once there are a number of examples of reading transition classifications, then Rdgai can use an LLM to automatically classify reading transitions for new variation units. Rdgai uses langchain, a framework for building pipelines with LLMs. It abstracts the particular model of LLM that is being used so that users can take advantage of different models by a variety of vendors. As new LLMs are released, they can be also used with Rdgai. Rdgai builds a prompt based on the description of each category and examples already found in the TEI XML. The number of examples of each category can be set with a command-line argument. Examples are prioritized to include ones with descriptions in the XML to give the prompt more information about the justification of each classification. To ensure a diversity of examples, a distance matrix is generated from the text of all readings in each category using the Levenshtein distance of the reading texts and then FasterPAM k-medoids clustering[12] is used to select a certain number of representative examples. The number of examples can be adjusted using the command line interface. Most of the initial part of the prompt template for a given document is the same for all queries which allows Rdgai to advantage of prompt caching by models which allow for it.

The LLM is asked to choose from one of the provided classification categories for the pairs of readings in a given variation unit. The resulting classification is added to the TEI XML as a relation element. Rdgai is specified in the resp attribute, indicating that Rdgai is the party responsible for this element of the file. This allows readers of the TEI XML to be able to track which classifications are manual and which are automatic. The LLM is also required to give a justification of its decision and this is included as a child desc element. This allows for interpretability and scrutiny of the results from Rdgai.

[12] Schubert and Rousseeuw, "Fast and Eager k-Medoids Clustering."

## Validation

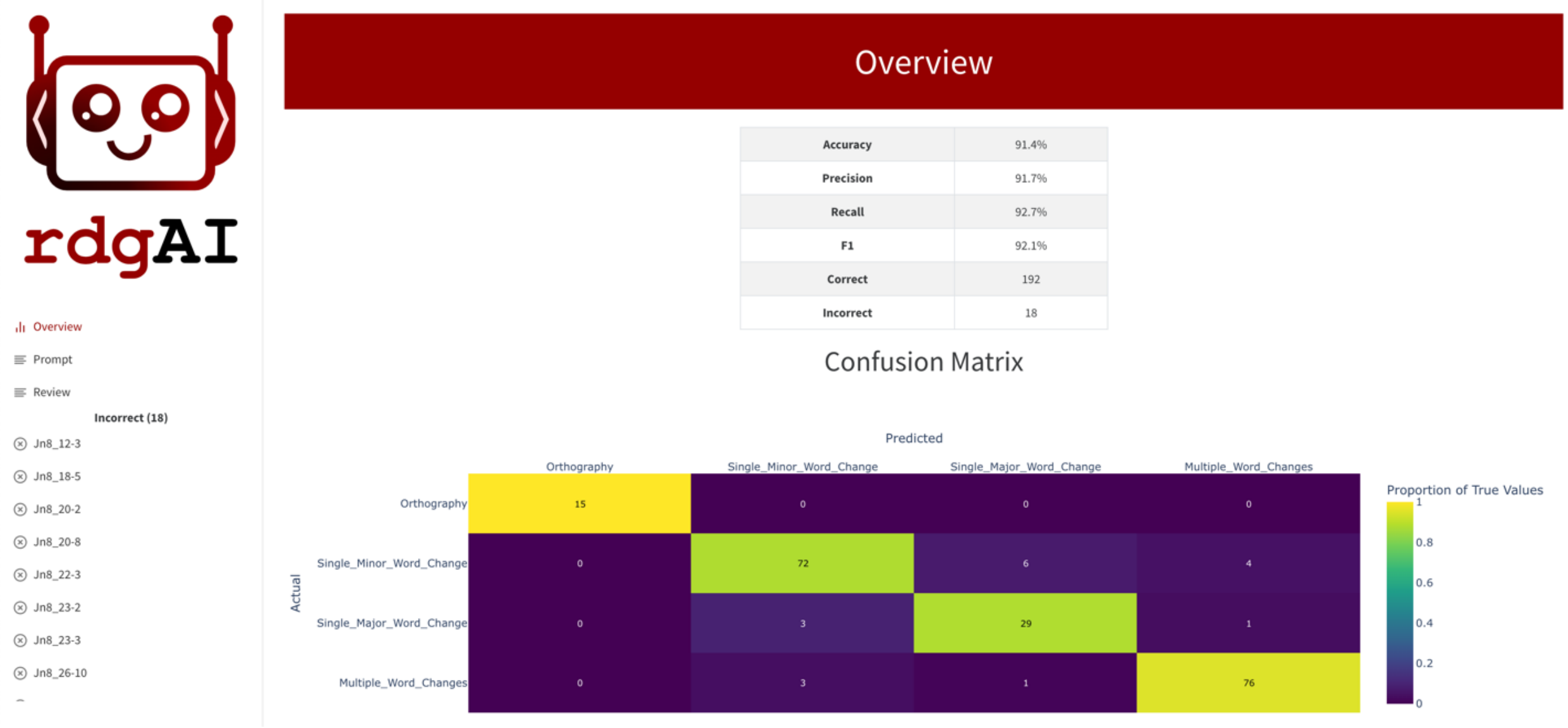

*Fig. 6. The Rdgai HTML report for evaluating performance.*

The accuracy of Rdgai is dependent on the type of text, the categories and their definitions and the LLM used. The accuracy needs to be validated on each document used with Rdgai. For this purpose, Rdgai comes with a validation tool which assigns a proportion of the manual annotations to be allowed for use in the prompt and the remainder are used as ground truth annotations for evaluating the results from Rdgai. It creates an HTML report (Fig. 6) showing the accuracy, macro precision, recall and F1 scores, a confusion matrix and a complete list of all the correct and incorrect classifications, showing the ground truth classification, the predicted category from Rdgai and the justification given. The report includes the base prompt including the representative example for each category. A textual version of the report is given to the LLM and it is asked to review the prompt with the category definitions and examples, based on the correct and incorrect results. The LLM then gives suggestions for clarifying the definitions of the categories and alerts the user to any inconsistencies in the ground truth annotations.

## Phylogenetic Analysis

Rdgai is designed to work with teiphy, a phylogenetic software package which takes a critical apparatus in TEI XML and converts it to the input file required for a number of different phylogenetic packages.[13] For using the classification of readings using Rdgai, teiphy can take the TEI XML output from Rdgai and convert it to an input file for the Bayesian phylogenetic application Beast 2. It creates the required substitution matrices for each of the variation units and adds default priors for the required rates.

[13] McCollum and Turnbull, “teiphy: A Python Package for Converting TEI XML Collations to NEXUS and Other Formats.”

## Evaluation on Test Dataset

To evaluate the results of Rdgai, we use a TEI XML collation of John 8:12–51 which collates the witnesses of ar[b]. This XML file is available in the Rdgai GitHub repository. Four reading transition categories are used:

- **Orthography**: Changes in spelling or diacritic marks. This includes any alterations to how words are written without affecting their meaning, such as standardizing spelling or correcting errors.
- **Single Minor Word Change**: An addition, omission or substitution to a single minor word or part of a word, including the substitution of a conjunction, pronoun, pronominal suffix, definite article, preposition or particle. This also includes an expansion or contraction of a word. It also includes changes to the verb form.
- **Single Major Word Change**: An addition, omission or substitution to a single word that is more significant than a minor word defined above.
- **Multiple Word Changes**: Changes across more than one word. Words can be major or minor.

The TEI XML file includes more than 800 manually annotated changes in readings. Half of these are available for use in the prompt and the remainder are used for evaluating the results of Rdgai.

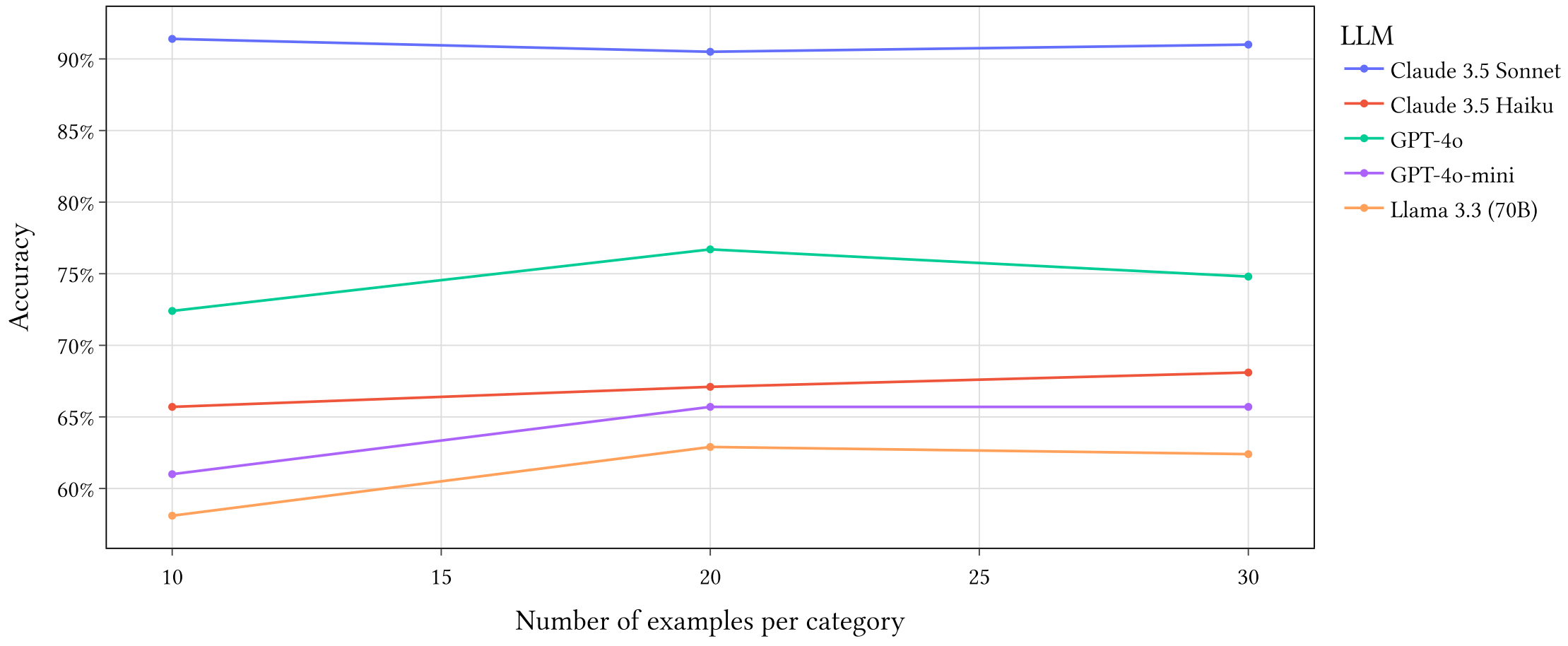

*Fig. 7. Accuracy versus the number of examples for each LLM.*

A number of LLMs were used: OpenAI's GPT-4o and GPT-4o-mini, Anthropic's Claude 3.5 Sonnet and Haiku, and Meta's 70B parameter Llama 3.3 model. For each, three sets of examples for each category were allowed: 10, 20 and 30. The accuracy results are shown in Fig. 7. The HTML reports for each experiment are available in the Rdgai GitHub repository.

The number of examples had little impact; having 20 or 30 examples per category did not greatly improve performance compared with merely 10. The choice of LLM had a far greater impact on accuracy than the number of examples per category. The best model was Claude 3.5 Sonnet, where the experiment with 10 examples achieved an accuracy of 91.4%. All Claude 3.5 Sonnet experiments achieved an accuracy of over 90%. Larger models such as Claude 3.5 Sonnet and GPT-4o performed better than their smaller counterparts (Claude 3.5 Haiku and GPT-4o-mini). Models without publicly available parameters (i.e. the Anthropic models and the Open AI models) performed better than Meta’s Llama with publicly accessible weights.

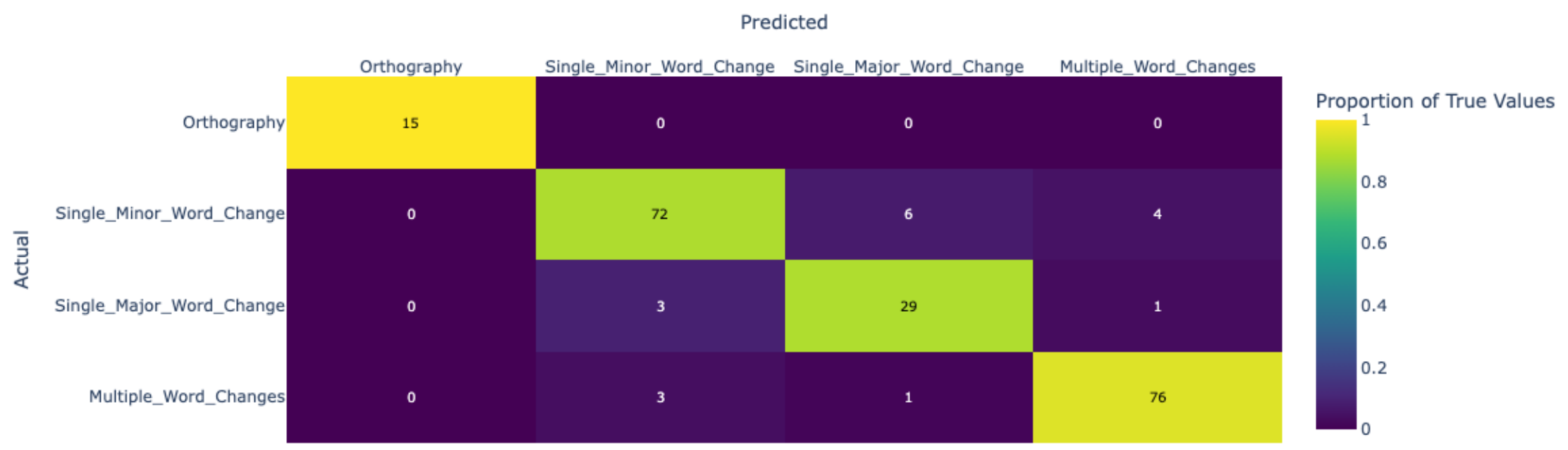

Fig. 8. The confusion matrix for the Claude 3.5 Sonnet model with 10 examples per category.

The confusion matrix for the best model is shown in Fig. 8. The largest source of error was six ‘Single Minor Word Changes’ being classified as ‘Single Major Word Changes’. For example, in Jn 8:18, the transition from ذلك لي to وهو لي, the following justification was given: ‘The demonstrative pronoun ذلك is changed to the pronoun وهو. While both are pronouns, this involves changing to a different word with a different root, not just modifying the form of the same word.’ Though changes to demonstrative pronouns were explicitly stated to be a minor word change in the description, there are no instances with ذلك in the examples.

There were also four instances where ‘Single Minor Word Changes’ were classified as ‘Multiple Word Changes’. For example, in Jn 8:26 the transition from اياه to فهو was given the justification: ‘This change involves two distinct words: انا is changed to ف and اياه is changed to هو. Since more than one word is affected, this must be classified as Multiple Word Changes. This is similar to the example “انا اياه → فهو" provided in the Multiple Word Changes category.’ Here the understanding of

These problems could potentially be addressed with more explicit descriptions in the TEI XML or by ensembling the results from multiple models into a single prediction. These avenues will be explored in future applications of Rdgai.

# Conclusion

This chapter presents Rdgai, a tool to automate the classification of changes in readings. Users take a critical apparatus in TEI XML format and manually annotate a number of changes in readings using the Rdgai GUI or by using Microsoft Excel. Rdgai then uses an LLM to classify other changes in readings and provide a justification for its decision. A validation tool is provided to evaluate the accuracy of the automated classifications including an HTML report with a confusion matrix, error analysis and suggestions to refine TEI XML input. On a test dataset based on a family of Arabic Gospels, Rdgai achieved an accuracy of greater than 90%. It is hoped that this tool will enable more sophisticated phylogenetic analysis of texts, accounting for diverse types of textual change in the results.

# Availability

The software is available under the Apache 2.0 license and is accessible on GitHub (https://github.com/rbturnbull/rdgai). The code has 100% testing coverage. Documentation is available at https://rbturnbull.github.io/rdgai/. The package can be downloaded using pip from the Python Package Index (https://pypi.org/project/rdgai/).

# Acknowledgments

This research was supported by The University of Melbourne's Research Computing Services. I thank Father Justin Sinaites for access to the manuscripts discussed in this article. I acknowledge the assistance of Joey McCollum, Emily Fitzgerald, Andrew Turner and Giulia Torello-Hill.